\documentclass{article}
\usepackage{template/spconf}
\usepackage{amsmath,graphicx,hyperref}
\usepackage{booktabs}
\usepackage{multicol,multirow}
\usepackage{caption}
\usepackage{cleveref}
\usepackage{todonotes}

\newcommand{\ie}{\textit{i.e.},\ }
\newcommand{\eg}{\textit{e.g.},\ }

\usepackage{glossaries}
\newacronym{ai}{AI}{artificial intelligence}
\newacronym{dl}{DL}{deep learning}
\newacronym{ml}{ML}{machine learning}
\newacronym{ssl}{SSL}{self-supervised learning}
\newacronym{ap}{AP}{average precision}
\newacronym{auroc}{AUROC}{area under the receiver operating characteristic curve}

\title{Representation-Based Data Quality Audits for Audio}

\name{%
\begin{tabular}{@{}c@{}}
Alvaro Gonzalez-Jimenez\sthanks{*Equal contribution}$^{1,3}$, Fabian Gröger$^{*1,2}$, Linda Wermelinger$^{1,2}$, \\ 
Andrin Bürli$^{4}$, Iason Kastanis$^{4}$, Simone Lionetti$^{1}$, Marc Pouly$^{1}$\thanks{We gratefully acknowledge the support of the Hasler Foundation grant 2024-10-17-192.}
\end{tabular}
}

\address{%
\begin{tabular}{@{}c@{}}
$^{1}$ Lucerne University of Applied Sciences and Arts \quad $^{2}$ University of Basel \\
$^{3}$ University Hospital of Basel \quad $^{4}$ CSEM
\end{tabular}
}

\begin{document}
%
\frenchspacing
\maketitle
\begin{abstract}
Data quality issues such as off-topic samples, near duplicates, and label errors often limit the performance of audio-based systems. 
This paper addresses these issues by adapting \mbox{SelfClean}, a representation-to-rank data auditing framework, from the image to the audio domain.
This approach leverages self-supervised audio representations to identify common data quality issues, creating ranked review lists that surface distinct issues within a single, unified process. 
The method is benchmarked on the ESC-50, GTZAN, and a proprietary industrial dataset, using both synthetic and naturally occurring corruptions. 
The results demonstrate that this framework achieves state-of-the-art ranking performance, often outperforming issue-specific baselines and enabling significant annotation savings by efficiently guiding human review.
\end{abstract}

\section{Introduction}
High‐stakes audio applications, from predictive maintenance and safety monitoring to large-scale media search, depend on data that is abundant and trustworthy~\cite{cote2024data, li2021_cleanml, northcutt_pervasive_2021}. 
Yet, in real-world settings, audio corpora often suffer from quality issues, such as off-topic samples (\ie inputs included by mistake), near duplicates (\ie redundant inputs from the same source), and label errors (\ie wrongly annotated samples).
These problems arise from fragmented data acquisition and annotation pipelines lacking standardization across sources, environments, and annotator expertise~\cite{sturm2012analysis,fonseca2019_fsdaudio,gemmeke2017_audioset}. 
Such inconsistencies not only degrade model accuracy but also obscure true generalization during evaluation~\cite{groger2024intrinsic}.
As a result, data quality auditing and maintenance are operational necessities not just academic concerns~\cite{northcutt_pervasive_2021,li2021_cleanml,sculley2015_hidden,polyzotis2017_production_ml}.

Hence, there has been growing interest in data cleaning methodologies applicable beyond structured data to modalities such as images, audio, or text. 
A prominent such framework is SelfClean~\cite{groger2024intrinsic}, a dataset auditing framework developed initially for images. 
The framework first learns \emph{intrinsic} representations---features learned directly from the target dataset itself. 
It then applies \emph{indicator} functions---metrics calculated on these representations---to score each sample for quality issues like off-topic content, near duplicates, and label errors. 
By presenting these issues in ranked lists, this design fits industrial workflows well, focusing on accelerating expert triage rather than automating decisions without oversight.

This paper explores how SelfClean can be transferred to the audio domain.
The key challenges in porting this framework are audio's temporal structure and modality-specific ambiguities (\eg ``off-topic'' can manifest as content, quality, or structural mismatch, and ``near duplicate'' can manifest at segment- or file-level under time shifts or recording differences). 
Furthermore, the representation-based approach of creating a dataset-specific latent space for SelfClean must be adapted to incorporate specific learning techniques that capture these unique challenges.

We tested the adapted SelfClean framework using modern audio encoders, benchmarking its ability to detect both synthetic and natural contamination.
Initially, we explored data-specific representations by training CLMR~\cite{spijkervet2021contrastive} from scratch on the target dataset, following the imaging-domain approach, but observed underperformance.
We then benchmarked several pre-trained, non-dataset-specific audio encoders such as CLMR~\cite {spijkervet2021contrastive}, CAV-MAE~\cite{gongcontrastive2023}, EAT~\cite{chen2024eat}, BEATs~\cite{chen2023beats}, and M2D~\cite{niizumi2024masked} trained on large-scale general audio corpora, which generalized well across contamination types.
Finally, we explored incorporating dataset context via LoRA~\cite{hu2022lora} adaptation with an unsupervised objective, mirroring SelfClean's original philosophy. 

Practically, we found that when adapting SelfClean to other domains, the representation part needs to be carefully examined and validated. 
In contrast, the indicator functions seamlessly integrate and generalize across modalities.

\textbf{Contributions.} 
\textbf{(1)} Porting SelfClean to audio by adapting the representation-learning stage to techniques that capture the unique properties of the domain and proposing a file-level aggregation strategy tailored for production. 
\textbf{(2)} Developing a production-ready implementation of SelfClean for audio, along with a comprehensive, reproducible evaluation protocol spanning synthetic corruptions, imperfect public benchmarks, and a private industrial dataset. 
\textbf{(3)} Conducting an extensive empirical study across state-of-the-art audio encoders, identifying where representation choice impacts performance and where SelfClean remains robust, along with operational lessons for deploying dataset audits in real pipelines.
\textbf{(4)} Utilizing the fraction of effort (FoE) metric to quantify review efficiency, linking ranking quality to operator time.

\section{Related work}

\textbf{Data quality.}
Data-centric research shows that label errors, spurious correlations, and other defects degrade training and evaluation, motivating systematic data auditing and curation before modeling \cite{cote2024data,northcutt_pervasive_2021}. 
In practice, such defects often arise from heterogeneous capture pipelines and loosely governed annotation workflows \cite{sculley2015_hidden}.
We follow the terminology of \cite{groger2024intrinsic}, classifying issues as off-topic samples, near duplicates, and label errors.

\textbf{Single-issue detection.}
Off-topic samples are typically addressed using outlier or novelty detection, often through unsupervised methods over learned embeddings \cite{chalapathy_anomaly_2019,zhao_pyod_2019,liu_isolation_2008}. 
Near-duplicate detection in audio has deep industrial roots, including fingerprinting systems such as Shazam, as well as alternative descriptors operating on spectrogram images \cite{wang_industrial_2003,burges_using_2005,williams_efficient_2017,drevo_worldveildejavu_2024}. 
Label error detection is often framed as reconciling model predictions and annotations \cite{northcutt_confident_2022,cleanlab2017cleanlab}.

\textbf{Multi-issue detection.}
Toolkits such as Cleanlab~\cite{cleanlab2017cleanlab} and FastDup~\cite{visuallayer2022fastdup} broaden the scope beyond a single defect type (\eg label errors or duplicates), but often depend on supervised task models, modality-specific heuristics, and lack proper academic validation. 
In contrast, SelfClean~\cite{groger2024intrinsic} introduced a  representation-to-rank paradigm.
Specifically, it first learns dataset-specific self-supervised embeddings and identifies off-topic samples via local structure in feature space, near-duplicates via proximity, and label errors via intra-class and extra-class ratios.

\textbf{Positioning.}
Prior audio cleaning approaches typically target one issue at a time and assume access to task-specific labels or models.
In contrast, building on SelfClean's ranking formulation, we propose an \emph{audio} variant that simultaneously detects off-topic samples, near duplicates, and label errors within a unified, lightweight framework designed for human-in-the-loop workflows in industrial pipelines.

\section{Method}
\label{sec:method}

\subsection{Problem definition}
Given a data collection, the goal of data quality auditing is to produce ranked lists for human review.
Specifically, one list per issue type: off-topic (OT), near-duplicate (ND), and label-error (LE). 
With the goal that samples appearing earlier in each ranking are more likely to correspond to the respective quality issue.

We evaluate these rankings using \gls*{auroc} and \gls*{ap}, metrics that reflect ranking quality, critical for accelerating manual quality control.

\subsection{Dataset}
We conduct experiments on three datasets. 
\textit{ESC-50}~\cite{piczak2015esc}: An environmental sound classification benchmark comprising 2,000 audio clips, each 5 seconds long, divided into 50 classes. 
These classes range from natural sounds (\eg rain, wind, animal vocalizations) to human activities and mechanical noises. 
\textit{GTZAN}~\cite{tzanetakis2002musical}: A dataset for music genre classification, consisting of 1,000 audio tracks, each 30 seconds long, evenly distributed across 10 genres. 
\textit{CSEM}: A proprietary dataset of membrane pump recordings across distinct hydraulic states, comprising 763 audio samples with a maximum length of 30 seconds, divided into four classes.

\subsection{Evaluation}
We perform two types of evaluations to benchmark different data quality detection methods.
First, we add synthetic noise to an assumed-to-be clean dataset, here ESC-50, and compare how well these artificial corruptions can be detected.
Second, we evaluate how well these methods can detect natural data quality issues by using datasets that contain known data quality issues, such as GTZAN, and a dedicated unseen industrial dataset with naturally occurring noise, followed by domain expert validation.
This diversity allows us to test the generalization and practical utility that extend beyond academic benchmarks.

\textbf{Synthetic evaluation.}
We construct a controlled protocol by injecting synthetic noise into ESC-50 to probe how well methods recover corrupted samples. 
For each issue type, we randomly select a subset of training samples with probability $\alpha \in \{0.05, 0.1, 0.2\}$.
Specifically, for creating \textit{near duplicates}, we sample uniformly from three transformations (additive perturbation, contiguous temporal crop, or a mixed/noisy variant).
For \textit{off-topic samples}, we replace samples by content drawn uniformly from a mixture of pure noise, length-matched audio from an external, unrelated corpus (here WMMS \cite{sayigh2016watkins}), or heavily noisy versions of the original.
For \textit{label errors}, the observed label of each selected sample is replaced by a different class chosen uniformly from the remaining classes. 
Together, these strategies capture typical imperfections encountered in practice, such as sensor malfunctions, annotation mistakes, and curation artifacts.

\textbf{Natural evaluation.} 
We assess performance on real, unmodified data by applying methods to two in-the-wild corpora: GTZAN and an industrial audio dataset from CSEM. 
For GTZAN, we evaluate against a curated set of documented issues, namely near-duplicates and label inconsistencies \cite{sturm2012analysis}. 
For the CSEM corpus, we rely on quantitative and qualitative evaluations by domain experts, conducting a detailed investigation of the produced rankings to assess the overall usability of the rankings.

\subsection{Audio embeddings}
We encode segments using various self-supervised trained backbones and aggregate segment embeddings into a file-level vector via mean pooling.
Specifically, we evaluate five audio representations.  
CLMR~\cite{spijkervet2021contrastive} uses contrastive learning on raw waveforms to learn representations.
CAV-MAE~\cite{gongcontrastive2023} combines contrastive audio-visual alignment with masked autoencoding, producing semantically rich and robust features.  
EAT~\cite{chen2024eat} is an efficient Transformer tailored for long audio sequences, offering a lightweight yet expressive representation.  
BEATs~\cite{chen2023beats} leverages a tokenizer to discretize audio features, emphasizing semantically meaningful tokens and facilitating more effective learning.  
M2D~\cite{niizumi2024masked} uses a self-supervised bootstrap framework on augmented spectrograms to capture latent audio representations during pre-training.  

\subsection{Implementation details}
We use public checkpoints for CLMR, CAV-MAE, EAT, BEATs, and M2D.
Specifically, CAV-MAE, EAT, BEATs, and M2D are pre-trained on AudioSet-2M, and CLMR is pre-trained on MagnaTagATune.
Inputs to all models are raw waveforms at 16 kHz. 
For the ``intrinsic'' encoder, we train CLMR on the target corpus for 1,300 epochs with standard parameters from \cite{spijkervet2021contrastive}.
Similarly, for LoRA, we adapt using the same self-supervised objective from CLMR, InfoNCE \cite{spijkervet2021contrastive}, and adapt Q, K, V, and fully connected parameters with rank=16, alpha scaling=48, lr=6e-5 for eight epochs.

\section{Results}

\begin{table}[t]
\centering
\caption{Performance of pre-trained representations on the ESC-50 dataset under synthetic noise strategies.}
\label{tab:res-representations}
\resizebox{\linewidth}{!}{%
\begin{tabular}{l l cc cc cc}
\toprule
& & \multicolumn{2}{c}{$\alpha=0.05$} & \multicolumn{2}{c}{$\alpha=0.1$} & \multicolumn{2}{c}{$\alpha=0.2$} \\
\cmidrule(lr){3-4} \cmidrule(lr){5-6} \cmidrule(lr){7-8}
\textbf{Issue} & \textbf{Model} & AUROC & AP & AUROC & AP & AUROC & AP \\
\midrule

\multirow{7}{*}{\parbox[c][4\baselineskip]{0.7cm}{\centering OT}} 
& CLMR~\cite{spijkervet2021contrastive} & 0.506 & 0.050 & 0.502 & 0.098 & 0.497 & 0.196 \\
& CAV-MAE~\cite{gongcontrastive2023} & 0.309 & 0.049 & 0.260 & 0.075 & 0.273 & 0.161 \\ 
& M2D~\cite{niizumi2024masked} & 0.689 & 0.074 & 0.510 & 0.095 & 0.373 & 0.159 \\
& EAT~\cite{chen2024eat} & 0.591 & 0.070 & 0.596 & 0.138 & 0.544 & 0.222 \\
& BEATs~\cite{chen2023beats} & \textbf{0.766} & \textbf{0.253} & \textbf{0.745} & 0.316 & \textbf{0.673} & \textbf{0.341} \\
 \cmidrule{2-8}
 & CLMR (SSL)   & 0.222 & 0.031 & 0.175 & 0.058 & 0.163 & 0.118 \\
 & BEATs (LoRA) & 0.724 & 0.202 & 0.743 & \textbf{0.330} & 0.653 & 0.313 \\
\midrule

\multirow{7}{*}{\parbox[c][4\baselineskip]{0.7cm}{\centering ND}} 
& CLMR~\cite{spijkervet2021contrastive} & 0.740 & 0.001 & 0.747 & 0.001 & 0.744 & 0.001 \\
& CAV-MAE~\cite{gongcontrastive2023} & 0.744 & 0.032 & 0.724 & 0.017 & 0.730 & 0.018 \\
& M2D~\cite{niizumi2024masked} & \textbf{0.992} & 0.606 & \textbf{0.993} & 0.587 & \textbf{0.993} & 0.617 \\
& EAT~\cite{chen2024eat} & 0.930 & 0.482 & 0.922 & 0.468 & 0.931 & 0.476 \\
& BEATs~\cite{chen2023beats} & 0.972 & 0.606 & 0.978 & \textbf{0.595} & 0.978 & \textbf{0.625} \\
 \cmidrule{2-8}
 & CLMR (SSL) & 0.911 & 0.400 & 0.888 & 0.393 & 0.898 & 0.384 \\
 & BEATs (LoRA) & 0.970 & \textbf{0.608} & 0.975 & 0.588 & 0.977 & 0.619 \\
\midrule

\multirow{7}{*}{\parbox[c][4\baselineskip]{0.7cm}{\centering LE}} 
& CLMR~\cite{spijkervet2021contrastive} & 0.477 & 0.049 & 0.484 & 0.094 & 0.492 & 0.197 \\
& CAV-MAE~\cite{gongcontrastive2023} & 0.721 & 0.222 & 0.693 & 0.299 & 0.658 & 0.387 \\
& M2D~\cite{niizumi2024masked} & \textbf{0.998} & \textbf{0.970} & \textbf{0.995} & \textbf{0.950} & \textbf{0.986} & \textbf{0.943} \\
& EAT~\cite{chen2024eat} & 0.969 & 0.668 & 0.969 & 0.759 & 0.954 & 0.793 \\
& BEATs~\cite{chen2023beats} & 0.996 & 0.927 & 0.992 & 0.908 & 0.980 & 0.903 \\
 \cmidrule{2-8}
 & CLMR (SSL) & 0.957 & 0.586 & 0.959 & 0.723 & 0.942 & 0.792 \\
 & BEATs (LoRA) & 0.997 & 0.932 & 0.992 & 0.915 & 0.978 & 0.903 \\
\bottomrule
\end{tabular}%
}
\end{table}

\textbf{Transferring SelfClean to audio.}
SelfClean couples \emph{intrinsic} self-supervision to learn dataset-specific embeddings and indicator functions in the representation space, thereby ranking off-topic samples, near duplicates, and label errors. 
Porting this to audio requires transferring both parts.
For self-supervision, we first reproduced the \emph{intrinsic} recipe by training a CLMR encoder directly on the target audio, named CLMR (SSL). 
On ESC-50 with synthetic corruptions (see Table~\ref{tab:res-representations}), intrinsic CLMR underperforms large, off-the-shelf audio encoders for OT detection. 
While it improves over generic CLMR for ND and LE, it still performs worse than modern pre-trained backbones across all issue types. 
This aligns with the known difficulty of learning strong audio representations from small corpora with short segments~\cite{cramer2019look}.

In contrast, SelfClean's indicator functions transfer unchanged and operate directly on audio embeddings aggregated at the file level. 
With pre-trained backbones, these indicators are highly effective.
Notably, BEATs embeddings outperform all others on average, though M2D representations are similarly competitive, except for OT detection, where they underperform. 

\textbf{Adaptation of representations with LoRA.}
To align with SelfClean's \emph{intrinsic} ethos while leveraging strong encoders, we adapted the best-performing pre-trained backbone (BEATs) using LoRA~\cite{hu2022lora} with a self-supervised InfoNCE~\cite{spijkervet2021contrastive} objective. 
In the evaluated synthetic setup, gains over frozen features were limited. 
LoRA adaptation showed minor or no improvements for ND and LE only at specific contamination rates and underperformed on OT detection.
In this synthetic setting, LoRA adaptation provided no significant benefit, likely due to the limited fine-tuning data and a mismatch between pre-training and fine-tuning objectives.

\textbf{SelfClean vs issue-specific detectors.}
We benchmarked SelfClean (using general BEATs features) against issue-specific detectors: Isolation Forest for OT and Confident Learning for LE (both in the same representation space), and Dejavu fingerprinting for ND on raw waveforms (see Table~\ref{tab:res-competing}). 
For OT, SelfClean is competitive at low corruption and significantly better as noise increases, with consistently higher AP. 
For ND, SelfClean outperforms fingerprinting across contamination rates.
For LE, SelfClean excels at low noise, while Confident Learning outperforms as label noise grows. 
Overall, SelfClean provides a strong, unified data quality detection method, and is the most robust when corruption rates are unknown or mixed.

\begin{table}[t]
\centering
\caption{Performance on the ESC-50 dataset under synthetic noise strategies of SelfClean and competing methods on the same representations.}
\label{tab:res-competing}
\resizebox{\linewidth}{!}{%
\begin{tabular}{l l cc cc cc}
\toprule
& & \multicolumn{2}{c}{$\alpha=0.05$} & \multicolumn{2}{c}{$\alpha=0.1$} & \multicolumn{2}{c}{$\alpha=0.2$} \\
\cmidrule(lr){3-4} \cmidrule(lr){5-6} \cmidrule(lr){7-8}
\textbf{Issue} & \textbf{Model} & AUROC & AP & AUROC & AP & AUROC & AP \\
\midrule

\multirow{2}{*}{\parbox[c][4\baselineskip]{0.7cm}{\centering OT}} 
 & IForest~\cite{liu_isolation_2008} & \textbf{0.791} & 0.212 & 0.676 & 0.177 & 0.406 & 0.188\\
 & SelfClean & 0.766 & \textbf{0.253} & \textbf{0.745} & \textbf{0.316} & \textbf{0.673} & \textbf{0.341} \\
\midrule

\multirow{2}{*}{\parbox[c][4\baselineskip]{0.7cm}{\centering ND}} 
 & Dejavu~\cite{drevo_worldveildejavu_2024} & 0.862 & 0.017 & 0.835 & 0.033 & 0.845 & 0.068 \\
 & SelfClean & \textbf{0.972} & \textbf{0.606} & \textbf{0.978} & \textbf{0.595} & \textbf{0.978} & \textbf{0.625} \\
\midrule

\multirow{2}{*}{\parbox[c][4\baselineskip]{0.7cm}{\centering LE}} 
 & CLearning~\cite{northcutt_confident_2022} & 0.994 & 0.884 & \textbf{0.994} & \textbf{0.951} & \textbf{0.993} & \textbf{0.973} \\
 & SelfClean & \textbf{0.996} & \textbf{0.927} & 0.992 & 0.908 & 0.980 & 0.903\\
\bottomrule
\end{tabular}%
}
\end{table}

\textbf{Natural contamination.}
To evaluate the behavior of detecting real-world noise, we utilized the GTZAN dataset and a private industrial dataset from CSEM. 
On GTZAN, SelfClean achieves near-perfect duplicate detection results (AUROC=1.000, AP=0.977), outperforming the audio-fingerprinting baseline Dejavu (AUROC=0.746, AP=0.003).
The almost perfect AP indicates that true duplicates are concentrated at the top of SelfClean's ranking, which is ideal for human-in-the-loop quality validation.
For label errors, Confident Learning slightly outperforms SelfClean (AUROC 0.791 vs 0.741, AP 0.207 vs 0.182).
This gap likely stems from subtle genre mislabels, where methods comparing model predictions to labels have an advantage.
In contrast, SelfClean's representation‐only indicators favor density and agreement cues rather than class‐specific contradictions.

On the CSEM dataset, SelfClean outperforms Dejavu in near-duplicate detection (AP=0.121 vs 0.063), although absolute performance remains low. 
This outcome is expected, as membrane pumps operate at high frequencies (60–70Hz), producing highly repetitive acoustic patterns in 30-second recordings. 
This performance gap relative to benchmarks underscores the challenge of generalizing to industrial applications.
Shorter sample windows or alternative STFT parameters may improve the results.
For off-topic sample detection, SelfClean (AP=0.328) outperforms Isolation Forest (AP=0.242). 
A qualitative inspection of false positives revealed that they primarily consist of recordings with inconsistent duration ($<\!30\text{s}$) or incidental background noises (\eg a laboratory door closing), which could reasonably be considered off-topic. 
Therefore, the reported score is a conservative estimate.
For label error detection, SelfClean (AP=0.750) clearly outperforms Confident Learning (AP=0.476), indicating that SelfClean's representation-based indicators are well-suited to identifying mislabeled samples in real-world industrial acoustic data.

\begin{figure}
    \centering
    \includegraphics[width=0.8\linewidth]{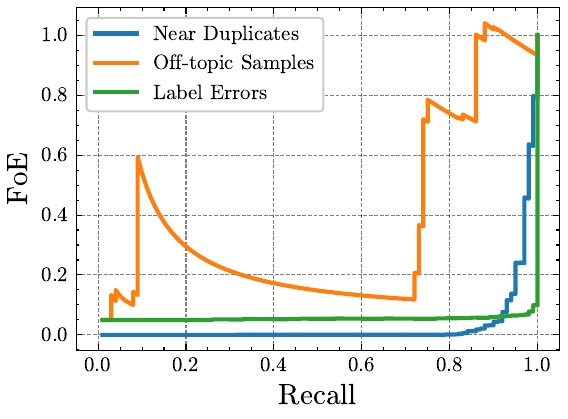}
    \caption{
        Ranking efficiency and annotation savings, where curves show $\mathrm{FoE}$ (fraction of effort, lower is better) as a function of recall for SelfClean's ranked lists on ESC-50 with synthetic corruption ($\alpha{=}0.05$) using BEATs embeddings.
        The steep curves for near duplicates and label errors indicate high review efficiency, enabling an operator to identify the majority of issues by inspecting a small fraction of the dataset.
    }
    \label{fig:FoE}
\end{figure}

\textbf{Annotation speed-ups.}
In human-in-the-loop workflows, the ranking itself determines throughput. 
Beyond AUROC/AP, we quantify efficiency as the \emph{fraction of effort (FoE)}~\cite{groger2024intrinsic}, which measures the proportion of the dataset an annotator must review to find a given fraction of issues compared to a random baseline, where $\mathrm{FoE}\!<\!1$ indicates time saved.
Figure~\ref{fig:FoE} shows FoE vs recall for the synthetic evaluation of SelfClean on BEATs embeddings with low contamination ($\alpha{=}0.05$). 
On average, SelfClean saves $97.1\%$ effort for near duplicates, $62.9\%$ for off-topic samples, and $94.6\%$ for label errors.
Equivalently, these translate into $34.2\times$ (ND), $2.69\times$ (OT), and $18.3\times$ (LE) speed-ups over an uninformed cleaning approach. 

\section{Conclusion}
This work demonstrates the successful adaptation of the SelfClean framework to the audio domain, establishing a powerful and unified methodology for data quality auditing. 
A key finding is that, unlike in the image domain, intrinsic self-supervised training on smaller target corpora is less effective.
Instead, leveraging large, pre-trained audio encoders like BEATs or M2D provides a robust foundation for data auditing ``out-of-the-box''.
Combined with SelfClean's indicator functions, these general-purpose representations achieve state-of-the-art performance in identifying off-topic samples, near-duplicates, and label errors. 
This process translates directly into significant operational efficiency, offering a speed-up of up to 34 times in annotation review for near duplicates. 
For practitioners, these findings provide a clear recommendation: frozen, off-the-shelf encoders offer an effective balance of performance and simplicity for real-world audio dataset maintenance. 
Future work will explore segment-level issue attribution and more advanced techniques for adapting general representations to specific dataset contexts.

\clearpage
\bibliographystyle{template/IEEEbib}
\bibliography{bibliography}

\begin{thebibliography}{10}

\bibitem{cote2024data}
Pierre-Olivier C{\^o}t{\'e}, Amin Nikanjam, et~al.,
\newblock ``Data cleaning and machine learning: a systematic literature review,''
\newblock {\em Automated Software Engineering}, 2024.

\bibitem{li2021_cleanml}
Peng Li, Xi~Rao, et~al.,
\newblock ``{CleanML: A Study for Evaluating the Impact of Data Cleaning on ML Classification Tasks},''
\newblock in {\em ICDE}, 2021.

\bibitem{northcutt_pervasive_2021}
Curtis~G. Northcutt, Anish Athalye, et~al.,
\newblock ``Pervasive label errors in test sets destabilize machine learning benchmarks,''
\newblock in {\em NeurIPS}, 2021.

\bibitem{sturm2012analysis}
Bob~L Sturm,
\newblock ``An analysis of the gtzan music genre dataset,''
\newblock in {\em ACM workshop on Music information retrieval with user-centered and multimodal strategies}, 2012.

\bibitem{fonseca2019_fsdaudio}
Eduardo Fonseca, Manoj Plakal, et~al.,
\newblock ``Learning sound event classifiers from web audio with noisy labels,''
\newblock in {\em ICASSP}, 2019.

\bibitem{gemmeke2017_audioset}
Jort~F. Gemmeke, Daniel P.~W. Ellis, et~al.,
\newblock ``Audio set: An ontology and human-labeled dataset for audio events,''
\newblock in {\em ICASSP}, 2017.

\bibitem{groger2024intrinsic}
Fabian Gr{\"o}ger, Simone Lionetti, et~al.,
\newblock ``Intrinsic self-supervision for data quality audits,''
\newblock {\em NeurIPS}, 2024.

\bibitem{sculley2015_hidden}
D.~Sculley, Gary Holt, et~al.,
\newblock ``Hidden technical debt in machine learning systems,''
\newblock in {\em NeurIPS}, 2015.

\bibitem{polyzotis2017_production_ml}
Neoklis Polyzotis, Sudip Roy, et~al.,
\newblock ``Data management challenges in production machine learning,''
\newblock in {\em ACM International Conference on Management of Data}, 2017.

\bibitem{spijkervet2021contrastive}
Janne Spijkervet and John~Ashley Burgoyne,
\newblock ``Contrastive learning of musical representations,''
\newblock {\em ISMIR}, 2021.

\bibitem{gongcontrastive2023}
Yuan Gong, Andrew Rouditchenko, et~al.,
\newblock ``Contrastive audio-visual masked autoencoder,''
\newblock in {\em ICLR}, 2023.

\bibitem{chen2024eat}
Wenxi Chen, Yuzhe Liang, et~al.,
\newblock ``Eat: Self-supervised pre-training with efficient audio transformer,''
\newblock in {\em IJCAI}, 2024.

\bibitem{chen2023beats}
Sanyuan Chen, Yu~Wu, et~al.,
\newblock ``Beats: audio pre-training with acoustic tokenizers,''
\newblock in {\em ICML}, 2023.

\bibitem{niizumi2024masked}
Daisuke Niizumi, Daiki Takeuchi, et~al.,
\newblock ``Masked modeling duo: Towards a universal audio pre-training framework,''
\newblock {\em TASLPRO}, 2024.

\bibitem{hu2022lora}
Edward~J Hu, Yelong Shen, et~al.,
\newblock ``Lora: Low-rank adaptation of large language models.,''
\newblock {\em ICLR}, 2022.

\bibitem{chalapathy_anomaly_2019}
Raghavendra Chalapathy, Aditya~Krishna Menon, et~al.,
\newblock ``Anomaly {Detection} using {One}-{Class} {Neural} {Networks},''
\newblock {\em arXiv}, 2019.

\bibitem{zhao_pyod_2019}
Yue Zhao, Zain Nasrullah, et~al.,
\newblock ``{PyOD}: A python toolbox for scalable outlier detection,''
\newblock {\em JMLR}, 2019.

\bibitem{liu_isolation_2008}
Fei~Tony Liu, Kai~Ming Ting, et~al.,
\newblock ``Isolation {Forest},''
\newblock in {\em {IEEE} {International} {Conference} on {Data} {Mining}}, 2008.

\bibitem{wang_industrial_2003}
Avery Wang et~al.,
\newblock ``An industrial strength audio search algorithm.,''
\newblock in {\em ISMIR}, 2003.

\bibitem{burges_using_2005}
C.J.C. Burges, D.~Plastina, et~al.,
\newblock ``Using audio fingerprinting for duplicate detection and thumbnail generation,''
\newblock in {\em ICASSP}, 2005.

\bibitem{williams_efficient_2017}
Dominic Williams, Akash Pooransingh, et~al.,
\newblock ``Efficient music identification using {ORB} descriptors of the spectrogram image,''
\newblock {\em {EURASIP} Journal on Audio, Speech, and Music Processing}, 2017.

\bibitem{drevo_worldveildejavu_2024}
Will Drevo,
\newblock ``Dejavu: open-source audio fingerprinting project,''
\newblock {\em Dejavu: open-source audio fingerprinting project}, 2014.

\bibitem{northcutt_confident_2022}
Curtis Northcutt, Lu~Jiang, et~al.,
\newblock ``Confident learning: Estimating uncertainty in dataset labels,''
\newblock {\em Journal of Artificial Intelligence Research}, 2021.

\bibitem{cleanlab2017cleanlab}
Jonas Mueller, Anish Athalye, et~al.,
\newblock ``Cleanlab,''
\newblock {\em GitHub. Note: https://github.com/cleanlab/cleanlab}, 2018.

\bibitem{visuallayer2022fastdup}
Amir Alush, Dickson Neoh, et~al.,
\newblock ``Fastdup,''
\newblock {\em GitHub. Note: https://github.com/visual-layer/fastdup}, 2022.

\bibitem{piczak2015esc}
Karol~J Piczak,
\newblock ``Esc: Dataset for environmental sound classification,''
\newblock in {\em ACM international conference on Multimedia}, 2015.

\bibitem{tzanetakis2002musical}
George Tzanetakis and Perry Cook,
\newblock ``Musical genre classification of audio signals,''
\newblock {\em IEEE Transactions on speech and audio processing}, 2002.

\bibitem{sayigh2016watkins}
Laela Sayigh, Mary~Ann Daher, et~al.,
\newblock ``The watkins marine mammal sound database: an online, freely accessible resource,''
\newblock in {\em Proceedings of Meetings on Acoustics}, 2016.

\bibitem{cramer2019look}
Aurora~Linh Cramer, Ho-Hsiang Wu, et~al.,
\newblock ``Look, listen, and learn more: Design choices for deep audio embeddings,''
\newblock in {\em ICASSP}, 2019.

\end{thebibliography}

\end{document}